\documentclass[journal,10pt]{IEEEtran}
\usepackage{amsmath}
\usepackage{bm}
\usepackage{algorithmic}
\usepackage{hyperref}
\usepackage{algorithm}
\usepackage{subfigure}
\usepackage{graphicx,epsfig,balance}
\usepackage{cite}
\usepackage{color}
\usepackage{multirow}
\usepackage{amsfonts,amssymb}
\usepackage{booktabs}
\usepackage{stfloats}
\usepackage{graphicx}
\begin{document}

\title{An Analysis of the Power Imbalance on the Uplink of Power-Domain NOMA}

\author{Shaokai Hu, Hao Huang, Guan Gui, \emph{Senior Member, IEEE}, and Hikmet Sari, \emph{Life Fellow, IEEE}\\ 
\IEEEcompsocitemizethanks{
\IEEEcompsocthanksitem S. Hu, H. Huang and G. Gui are with the College of Telecommunications and Information Engineering, Nanjing University of Posts and Telecommunications, Nanjing 210003, China (e-mail: 1221014234@njupt.edu.cn, 1017010502@njupt.edu.cn, guiguan@njupt.edu.cn).
\IEEEcompsocthanksitem H. Sari is with the College of Telecommunications and Information Engineering, Nanjing University of Posts and Telecommunications, Nanjing 210003, China, and also with the Sequans Communications, 15-55 Boulevard Charles de Gaulle, 92700 Colombes, France (e-mail: hsari@ieee.org).
}}

\markboth{submitted to IEEE Communications Letters}{}
\maketitle

\begin{abstract}
This paper analyzes the power imbalance factor on the uplink of a 2-user Power-domain NOMA system and reveals that the minimum value of the average error probability is achieved when the user signals are perfectly balanced in terms of power as in Multi-User MIMO with power control. The analytic result is obtained by analyzing the pairwise error probability and exploiting a symmetry property of the error events. This result is supported by computer simulations using the QPSK and 16QAM signal formats and uncorrelated Rayleigh fading channels. This finding leads to the questioning of the basic philosophy of Power-domain NOMA and suggests that the best strategy for uncorrelated channels is to perfectly balance the average signal powers received from the users and to use a maximum likelihood receiver for their detection.
\end{abstract}


\IEEEpeerreviewmaketitle

\section{Introduction}
\label{sec1}
Non-orthogonal multiple access (NOMA) is currently viewed as one of the key physical layer (PHY) technologies to meet the requirements of 5G network evolutions as well as those of Beyond 5G wireless networks. Since 2013, there has been a vast amount of literature on the subject, and using information theoretic arguments it was shown that NOMA can increase capacity over conventional orthogonal multiple access (OMA). Also, while other NOMA schemes have been proposed and investigated, the vast majority of the NOMA literature has been devoted to the so-called Power-domain NOMA (PD-NOMA), which is based on imposing a power imbalance between user signals and detecting these signals using a successive interference cancellation (SIC) receiver (see, e.g., \cite{YS,ZD,LD,ZDING,XLEI,SY,MS}). The specialized literature basically divides NOMA into two major categories, one being Power-domain NOMA and the other Code-domain NOMA, although there are approaches in some other domains (see, e.g., \cite{HSLETTER}). It is also worth mentioning the recently revived NOMA-2000 concept \cite{HSMAGAZINE,HSA,AME,AAX} that is based on using two sets of orthogonal signal waveforms \cite{IC,MV}. This technique does not require any power imbalance between user signals and falls in the Code-domain NOMA category due to the fact that the signal waveforms in one of the two sets are spread in time or in frequency.

Well before the recent surge of literature on NOMA during the past decade, the concept of multiple-input multiple-output (MIMO) was generalized to Multi-User MIMO (MU-MIMO), where the multiple antennas on the user side are not employed by the same user, but instead they correspond to multiple users \cite{QH,AM,SK,XC,BF}. For example, considering the uplink of a cellular system, two users with a single antenna each that are transmitting signals to the base station (BS) equipped with multiple antennas form an MU-MIMO system. Note that MU-MIMO is also known as Virtual MIMO as it can be seen from the titles of \cite{SK,XC,BF}. Naturally, the question arises as to what relationship PD-NOMA has with the MU-MIMO concept and how it compares to it in terms of performance. When the users in MU-MIMO share the same time and frequency resources simultaneously, the system becomes a NOMA scheme, and the only difference from PD-NOMA in this case is that user signals inherently have different powers in PD-NOMA while no power difference is involved in MU-MIMO. Therefore, this type of MU-MIMO can be referred to as Power-Balanced NOMA or Equal-Power NOMA.

In this paper, focusing on the uplink, we give a unified presentation of PD-NOMA and MU-MIMO by introducing a power imbalance parameter in the system model and we optimize this parameter to minimize the average bit error probability (ABEP) for a given total transmit power by the users. In the considered system scenario, the channels are uncorrelated Rayleigh fading channels, which is a typical assumption for the cellular uplink. The study reveals a very important result. Specifically, it was found that the minimum value of ABEP is achieved with zero power imbalance, i.e., when equal average powers are received by the BS from each of the users. This means that the power imbalance, which is the basic principle of PD-NOMA, is actually undesirable when the fading channels are uncorrelated.

The paper is organized as follows. In Section \ref{sec2}, we briefly recall the principle of PD-NOMA and MU-MIMO and give a unified system model which covers both of these techniques. Next, in Section \ref{sec3}, we optimize the power imbalance parameter in this model to minimize the ABEP, and we show that the optimum corresponds to the case of perfect power balance, as in MU-MIMO. In Section \ref{sec4}, we report the results of computer simulations confirming the theoretical findings of Section \ref{sec3}. Finally, Section \ref{sec5} summarizes our conclusions and points out some future research directions.

\section{Power Domain NOMA and Multi-User MIMO}
\label{sec2}
\subsection{Power-Domain NOMA}
\begin{figure}[htbp]
  \centering
  \includegraphics[width=3.0 in]{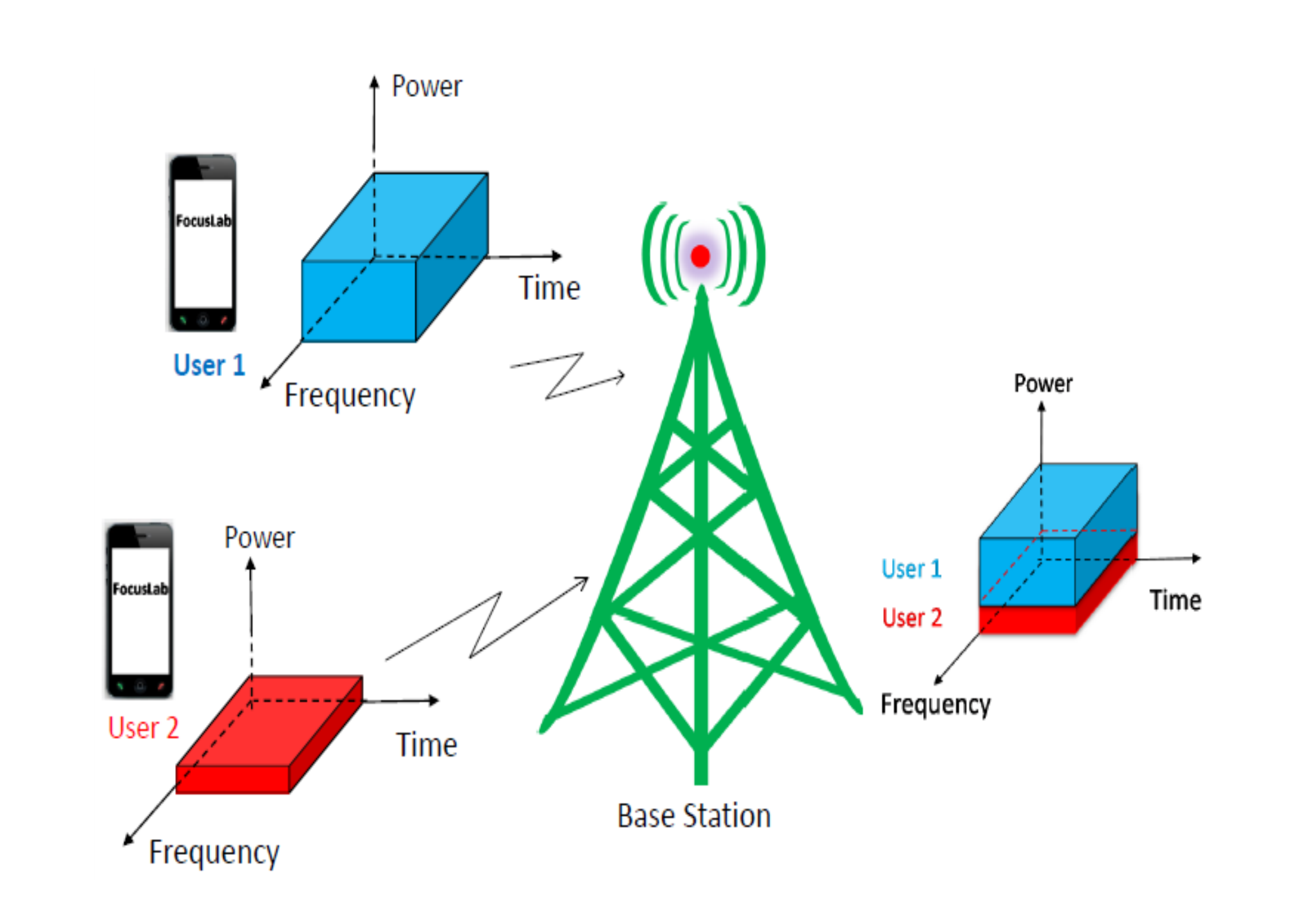}
  \caption{Illustration of Power-domain NOMA uplink with two users.}
  \label{Fig:Illustration of PD NOMA}
\end{figure}

The principle of PD-NOMA is to transmit user signals on the same time and frequency resource blocks by assigning different powers to them. On the receiver side, a SIC receiver is employed to detect the user signals. Fig. \ref{Fig:Illustration of PD NOMA} shows the concept of a 2-user PD-NOMA uplink. User 1 in this figure transmits a high signal power shown in blue, and User 2 transmits a low signal power shown in red. Assuming that the path loss is the same for both users, these signals arrive to the BS with the same power imbalance, and the BS detects them using the well-known SIC concept. Specifically, it first detects the strong User 1 signal, and then, it subtracts the interference of this signal on the weak User 2 signal in order to detect the latter signal.

The concept of superposing a strong user signal and a weak user signal and their detection using SIC has long been described in textbooks (see, e.g., \cite{DT}) in order to show that orthogonal multiple access is not optimal and that a higher capacity can be achieved by going to non-orthogonal multiple access. The NOMA literature, which basically started in 2013 directly followed this concept, and the resulting scheme was dubbed PD-NOMA as it assigns different powers to users. Note that although the SIC receiver is the main receiver which appears in the literature, receivers based on maximum-likelihood (ML) detection were also investigated in some recent papers \cite{JS,HS}, where higher performance was reported at the expense of an increased complexity.

\subsection{Multi-User MIMO}

Multi-User MIMO is the terminology given to a MIMO system when the multiple antennas on the user side do not correspond to a single user. For example, a cellular system in which a BS equipped with multiple antennas communicating with a number of single-antenna users forms an MU-MIMO system. Note that an MU-MIMO system is a simple OMA scheme if the users do not make simultaneous use of the same time and frequency resources, but when these resources are shared simultaneously it becomes a NOMA scheme, and this is what we consider here. Also note that with respect to conventional point-to-point MIMO links, MU-MIMO is what orthogonal frequency-division multiple access (OFDMA) is to orthogonal frequency-division multiplexing (OFDM). While OFDM and MIMO refer to point-to-point transmission links, OFDMA and MU-MIMO designate multiple access techniques based on the same principles. As in the PD-NOMA outlined in the previous subsection, here too we will focus on the uplink of a 2-user MU-MIMO system. In fact, Fig. 1 can also be used to describe this MU-MIMO system if the user signals have equal power and the BS is equipped with multiple antennas, which is always the case in state-of-the-art cellular networks. Therefore, the MU-MIMO we consider here is an equal-power NOMA system, in which the SIC receiver is not appropriate for signal detection. For both PD-NOMA and MU-MIMO, the optimum receiver is in fact the ML receiver, which makes its decisions by minimizing the Euclidean distance from the received noisy signal. In PD-NOMA with a large power imbalance between user signals, the SIC receiver essentially provides the ML detector performance, but this concept is not applicable to a power-balanced MU-MIMO system, because detection of one of the signals in the presence of interference from the other signal will lead to an excessive error rate.

\subsection{Unified System Model}
We now give a unified simple model for the uplinks in 2-user PD-NOMA and MU-MIMO systems with one antenna on the user side and two antennas on the BS side. Omitting the time index (which is not needed), the signals received by the first antenna and the second antenna of the BS are respectively given by:
\begin{align}
\label{eq1}
r_{1}=\sqrt{\alpha }h_{11}x_{1}+\sqrt{1-\alpha }h_{12}x_{2}+w_{1}
\end{align}
and
\begin{align}
\label{eq2}
r_{2}=\sqrt{\alpha }h_{21}x_{1}+\sqrt{1-\alpha }h_{22}x_{2}+w_{2}
\end{align}
where $x_1$ and $x_2$ are the symbols transmitted by the first and the second user, respectively, $\alpha$ is the power imbalance factor ($1/2\le\alpha<1$) with $\alpha=1/2$ corresponding to MU-MIMO, $h_{ij}$ designates the response of the channel between user $j$ and receive antenna $i$, and $w_1$ and $w_2$ are the additive white Gaussian noise (AWGN) terms. The channels are assumed to be unity-variance uncorrelated Rayleigh fading channels, and with this assumption the power imbalance at the receiver is identical to that imposed at the user side. In practice the users are randomly distributed within the cell, and the signal of each user is subjected to a path loss that depends on its distance to the BS. But in these situations, an appropriate power control can be used to impose a power imbalance factor $\alpha$ at the receiver, and the model above remains valid.

Let us combine equations (\ref{eq1}) and (\ref{eq2}) and write the vector equation:
\begin{align}
\label{eq3}
R=HX+W
\end{align}
where $R=\begin{pmatrix}
r_{1} \\r_{2}
\end{pmatrix}, H=\begin{pmatrix}
 h_{11}&h_{12}  \\
 h_{21}&h_{22}  \\
\end{pmatrix}, X=\begin{pmatrix}
\sqrt{\alpha }x_{1} \\\sqrt{1-\alpha }x_{2}
\end{pmatrix}$, and $W=\begin{pmatrix}
w_{1} \\w_{2}
\end{pmatrix}$.

The ML receiver makes its decisions by minimizing the Euclidean distance metric $||R-HX||^2$ over all values of the symbol vector $X$. For a constellation size of $M$, this involves the computation of  $M^2$ metrics and their comparisons in order to find the minimum value. Obviously, the ML receiver complexity is higher than that of the SIC receiver, which only involves the computation of 2$M$ metrics \cite{HS}.

\section{Optimizing the Power Imbalance}
\label{sec3}
The problem we address now is to determine the optimum power imbalance factor $\alpha$ leading to the smallest ABEP for a given total transmit power by the two users. The ABEP is usually evaluated by first evaluating the pairwise error probability (PEP), which corresponds to detecting a different codeword (symbol vector) from the transmitted one. Once the PEP is evaluated for all error events, it is used to derive the ABEP through weighing, summing, and averaging over all codewords. Note that an error event occurs when at least one of the symbols in the transmitted codeword is detected in error. In the case at hand, the error event implies that a symbol error occurs for at least one of the users. The ABEP can be upper bounded using the well-known union bound:
\begin{align}
\label{eq4}
ABEP\leqslant \frac{1}{M^{2}}\sum _{X}\sum _{\hat{X}\neq X}\frac{N(X,\hat{X})}{2\log_{2}(M)}P(X\to\hat{X} )
\end{align}
where $P(X\to\hat{X})$ denotes the PEP corresponding to the transmission of codeword $X$ and detection of codeword $\hat{X}\neq X$, and $N(X,\hat{X})$ denotes the number of bits in error corresponding to that error event. The parameter $M$ is the number of constellation points, and consequently the denominator $2\log_{2}(M)$ in these sums is the number of bits per transmitted codeword. The sum indexed $\hat{X}\neq X$ represents the ABEP conditional on the transmission of a codeword $X$. Finally, the sum indexed $X$ and the division by $M^2$ are for averaging the conditional bit error probabilities with respect to the transmitted codewords. Note that Eq. (\ref{eq4}) is written assuming that both users transmit symbols from the same constellation, and this assumption is made in the following analysis.

Using the pairwise error probability, we will now use a simple technique to demonstrate that the optimum NOMA scheme corresponding to the system model in Subsection II.C is in fact the NOMA scheme that is perfectly balanced in terms of transmit power. An upper bound on the PEP for spatial multiplexing type 2x2-MIMO systems operating on uncorrelated Rayleigh fading channels is given in [20, p. 79] as:
\begin{align}
\label{eq5}
P(X\to \hat{X})\leq \begin{bmatrix}
\frac{1}{1+\frac{1}{4N_{0}} \begin{Vmatrix}
X-\hat{X}\end{Vmatrix}^{2}}\end{bmatrix}^{2}
\end{align}

In this equation, $1/N_{0}$ is the signal-to-noise ratio (the bit energy being normalized by one), and referring back to the definition of $X$, the vector $X-\hat{X}$ is given by:
\begin{align}
\label{eq6}
X-\hat{X}=\begin{pmatrix}
\sqrt{\alpha }(x_{1}-\hat{x}_{1}) \\\sqrt{1-\alpha }(x_{2}-\hat{x}_{2})
\end{pmatrix}
\end{align}
where $\hat{x}_{1}$ (resp. $\hat{x}_{2}$) denotes the decision made on symbol $x_{1}$ (resp. $x_{2}$).

Let us consider now two symmetric error events $E_1$ and $E_2$ defined as $(x_1-\hat{x}_{1}=u,x_2-\hat{x}_{2}=v )$ and $(x_1-\hat{x}_{1}=v,x_2-\hat{x}_{2}=u )$, respectively, and without any loss of generality, assume $|u|>|v|$ . The squared Euclidean norm which appears in the denominator on the right-hand side of eqn. (\ref{eq5}) can be written as:
\begin{align}
\label{eq7}
\begin{Vmatrix}
X-\hat{X}\end{Vmatrix}_{E_{1}}^{2}=\alpha \begin{vmatrix}
u\end{vmatrix}^{2}+(1-\alpha)\begin{vmatrix}
v\end{vmatrix}^{2}
\end{align}
for error event $E_1$, and

\begin{align}
\label{eq8}
\begin{Vmatrix}
X-\hat{X}\end{Vmatrix}_{E_{2}}^{2}=\alpha \begin{vmatrix}
v\end{vmatrix}^{2}+(1-\alpha)\begin{vmatrix}
u\end{vmatrix}^{2}
\end{align}
for error event $E_2$.

Note that for $\alpha=1/2$, which corresponds to power-balanced NOMA (or MU-MIMO), we have:
\begin{align}
\label{eq9}
\begin{Vmatrix}
X-\hat{X}\end{Vmatrix}_{\alpha =1/2}^{2}=\frac{1}{2} (\begin{vmatrix}
u\end{vmatrix}^{2}+\begin{vmatrix}
v\end{vmatrix}^{2})
\end{align}
in both error events.

Using (\ref{eq7})-(\ref{eq9}), we can write:
\begin{align}
\begin{split}
\label{eq10}
&\begin{Vmatrix}
X-\hat{X}\end{Vmatrix}_{E_{1}}^{2}-\begin{Vmatrix}
X-\hat{X}\end{Vmatrix}_{\alpha=1/2}^{2} \\&=\alpha \begin{vmatrix}
u\end{vmatrix}^{2}+(1-\alpha)\begin{vmatrix}
v\end{vmatrix}^{2}-\frac{1}{2} (\begin{vmatrix}
u\end{vmatrix}^{2}+\begin{vmatrix}
v\end{vmatrix}^{2}) \\&=(\alpha-1/2)\begin{vmatrix}
u\end{vmatrix}^{2}+(1-\alpha-1/2)\begin{vmatrix}
v\end{vmatrix}^{2} \\&=(\alpha-1/2)(\begin{vmatrix}
u\end{vmatrix}^{2}-\begin{vmatrix}
v\end{vmatrix}^{2})>0
\end{split}
\end{align}
and
\begin{align}
\begin{split}
\label{eq11}
&\begin{Vmatrix}
X-\hat{X}\end{Vmatrix}_{E_{2}}^{2}-\begin{Vmatrix}
X-\hat{X}\end{Vmatrix}_{\alpha=1/2}^{2} \\&=\alpha \begin{vmatrix}
v\end{vmatrix}^{2}+(1-\alpha)\begin{vmatrix}
u\end{vmatrix}^{2}-\frac{1}{2} (\begin{vmatrix}
u\end{vmatrix}^{2}+\begin{vmatrix}
v\end{vmatrix}^{2}) \\&=(\alpha-1/2)\begin{vmatrix}
v\end{vmatrix}^{2}+(1-\alpha-1/2)\begin{vmatrix}
u\end{vmatrix}^{2} \\&=-(\alpha-1/2)(\begin{vmatrix}
u\end{vmatrix}^{2}-\begin{vmatrix}
v\end{vmatrix}^{2})<0
\end{split}
\end{align}

Referring back to the PEP upper bound in (\ref{eq5}), the squared distances $||X-\hat{X}||_{E_{1}}^{2}$and$||X-\hat{X}||_{E_{2}}^{2}$corresponding to the two considered error events are symmetric with respect to the squared distance $||X-\hat{X}||_{\alpha=1/2}^{2}$, and exploiting the Gaussian-like shape of the $ \lceil \frac{1}{1+z^{2}}\rceil^{2}$ function, we can write:
\begin{align}
\begin{split}
\label{eq12}
&2\begin{bmatrix}
\frac{1}{1+\frac{1}{4N_{0}}\begin{Vmatrix}
X-\hat{X}\end{Vmatrix}_{\alpha=1/2 }^{2}}\end{bmatrix}^{2}\\
&< \begin{bmatrix}
\frac{1}{1+\frac{1}{4N_{0}}\begin{Vmatrix}
X-\hat{X}\end{Vmatrix}_{E_{1}}^{2}}\end{bmatrix}^{2}+\begin{bmatrix}
\frac{1}{1+\frac{1}{4N_{0}}\begin{Vmatrix}
X-\hat{X}\end{Vmatrix}_{E_{2}}^{2}}\end{bmatrix}^{2}
\end{split}
\end{align}

This means that with  $\alpha\neq1/2$ corresponding to PD-NOMA, while the error event $E_1$ leads to a lower PEP than in the power-balanced case, error event $E_2$ leads to a higher PEP, and the sum of the two PEPs is smaller in the amplitude-balanced case. Note that this property holds for all pairs of symmetric error events in which $|u|^2\neq|v|^2$, and the PEP corresponding to error events with $|u|^2=|v|^2$ is independent of the power imbalance factor $\alpha$. Consequently, by averaging the PEP over all error events determined by the signal constellation, we find that the smallest average error probability is achieved when the NOMA scheme is perfectly balanced in terms of power.

We will now illustrate this property of the error events using the QPSK signal constellation assuming a power imbalance factor $\alpha=0.9$ (which corresponds to a power imbalance of 9.5 dB) and $1/N_0=100$, which corresponds to an SNR of 20 dB. Note that the PEP in this constellation is independent of the transmitted codeword, and therefore we can assume that the transmitted codeword is ($x_1=1+j,x_2=1+j$ ), and examine the 15 possible error events corresponding to the transmission of this codeword. Table \ref{TABLE1} shows the value of $||X-\hat{X}||^2$ as well as the PEP for $\alpha=0.5$ and $\alpha=0.9$ corresponding to the 15 error events, denoted ($E_1,E_2,...,E_{15}$).
\begin{table}
  \caption{Comparison of Power-Balanced NOMA ($\alpha=0.5$) and PD-NOMA with $\alpha=0.9$ in terms of PEP corresponding to different error events.}
  \Large
  \begin{center}
  \resizebox{\linewidth}{!}{
  \renewcommand\arraystretch{0.8}
  \begin{tabular}{cccccc}
  \toprule
  Error Event&$x_{1}-\hat{x}_{1}$&$x_{2}-\hat{x}_{2}$&$||X-\hat{X}||^{2}$&PEP($\alpha=0.5$)&PEP($\alpha=0.9$)\\
  \midrule
  $E_{1}$&$2$&$0$&$4\alpha$&$3.84\times10^{-4}$&$1.2\times10^{-4}$\\
  $E_{2}$&$0$&$2$&$4(1-\alpha)$&$3.84\times10^{-4}$&$\textcolor{red}{8.1\times10^{-3}}$\\
  $E_{3}$&$2j$&$0$&$4\alpha$&$3.84\times10^{-4}$&$1.2\times10^{-4}$\\
  $E_{4}$&$0$&$2j$&$4(1-\alpha)$&$3.84\times10^{-4}$&$\textcolor{red}{8.1\times10^{-3}}$\\
  $E_{5}$&$2$&$2$&$4$&$10^{-4}$&$10^{-4}$\\
  $E_{6}$&$2$&$2j$&$4$&$10^{-4}$&$10^{-4}$\\
  $E_{7}$&$2j$&$2$&$4$&$10^{-4}$&$10^{-4}$\\
  $E_{8}$&$2j$&$2j$&$4$&$10^{-4}$&$10^{-4}$\\
  $E_{9}$&$2+2j$&$0$&$8\alpha$&$10^{-4}$&$3\times10^{-5}$\\
  $E_{10}$&$0$&$2+2j$&$8(1-\alpha)$&$10^{-4}$&$\textcolor{red}{2.3\times10^{-3}}$\\
  $E_{11}$&$2+2j$&$2$&$4(1+\alpha)$&$4.3\times10^{-5}$&$2.7\times10^{-5}$\\
  $E_{12}$&$2$&$2+2j$&$4(2-\alpha)$&$4.3\times10^{-5}$&$8.1\times10^{-5}$\\
  $E_{13}$&$2+2j$&$2j$&$4(1+\alpha)$&$4.3\times10^{-5}$&$2.7\times10^{-5}$\\
  $E_{14}$&$2j$&$2+2j$&$4(2-\alpha)$&$4.3\times10^{-5}$&$8.1\times10^{-5}$\\
  $E_{15}$&$2+2j$&$2+2j$&$8$&$2.5\times10^{-5}$&$2.5\times10^{-5}$\\
  \bottomrule
  \end{tabular}}
  \end{center}
  \label{TABLE1}
\end{table}

The table shows that error event $E_1$ leads to a slightly higher PEP with $\alpha=0.5$, but with $\alpha=0.9$ the symmetric error event $E_2$ leads to a PEP that is higher by more than an order of magnitude. The same observation holds for the error events ($E_3,E_4$). Next, error events ($E_5,E_6,E_7,E_8$) lead to the same PEP of $10^{-4}$ for both values of $\alpha$. Proceeding further, while error event $E_9$ leads to a higher PEP for $\alpha=0.5$, its symmetric event $E_{10}$ gives a PEP that is more than an order of magnitude higher for $\alpha=0.9$. Finally, the error events ($E_{11},E_{12},E_{13},E_{14},E_{15}$) lead to PEP values in the range of $10^{-5}-10^{-4}$, with a small difference between the two values of $\alpha$. To get an upper bound on the ABEP, we compute the sum of the 15 PEPs after weighing them by the corresponding number of bits and we divide it by 4 because each codeword in QPSK carries 4 bits (2 bits per user). By doing this, we get an ABEP upper bound of $8\times10^{-4}$ for PD-NOMA with $\alpha=0.5$ and of $5\times10^{-3}$ for PD-NOMA with $\alpha=0.9$, the latter being dominated by the PEP values indicated in red on the last column of the table, which correspond to error events ($E_{2},E_{4},E_{10}$). This result shows that compared to $\alpha=0.5$, the power imbalance factor $\alpha=0.9$ increases the ABEP by almost an order of magnitude. Higher values of $\alpha$ incurred higher BER degradations that are not shown in Table \ref{TABLE1}.

\section{Simulation Results}
\label{sec4}
Using the QPSK and 16QAM signal formats, a simulation study was performed to evaluate the influence of the power imbalance factor on the bit error rate (BER) of PD-NOMA and confirm the theoretical findings reported in the previous section. Following the mathematical description of Section \ref{sec3}, we considered an uplink with two users transmitting toward a BS on two uncorrelated Rayleigh fading channels. Also as described in Section \ref{sec3}, the receiver employs ML detection and assumes that the channel state information (CSI) is perfectly known.

\begin{figure}[htbp]
  \centering
  \includegraphics[width=3.0 in]{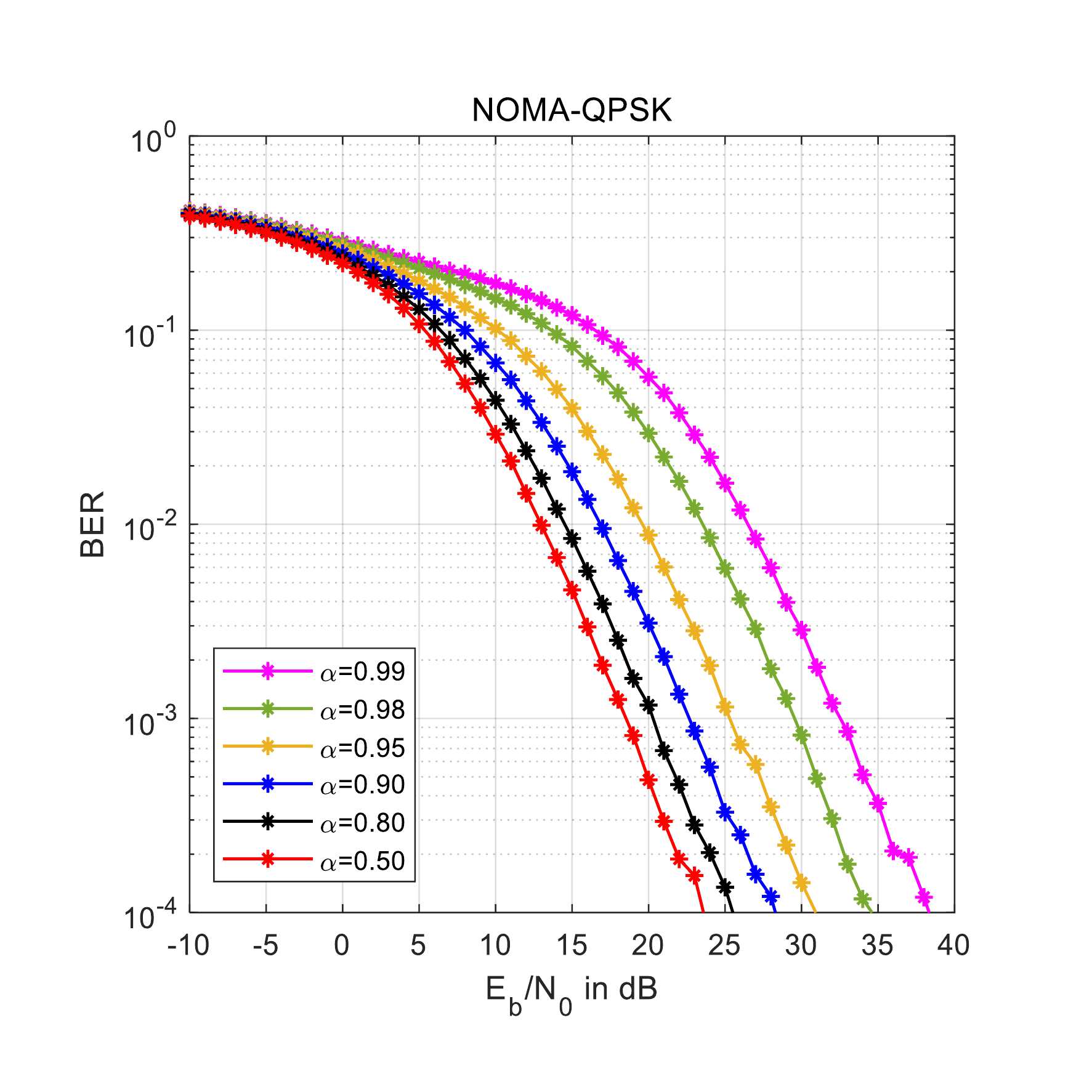}
  \caption{BER performance of PD-NOMA with QPSK and different values of the power imbalance.}
  \label{Fig:QPSK}
\end{figure}

\begin{figure}[htbp]
  \centering
  \includegraphics[width=3.0 in]{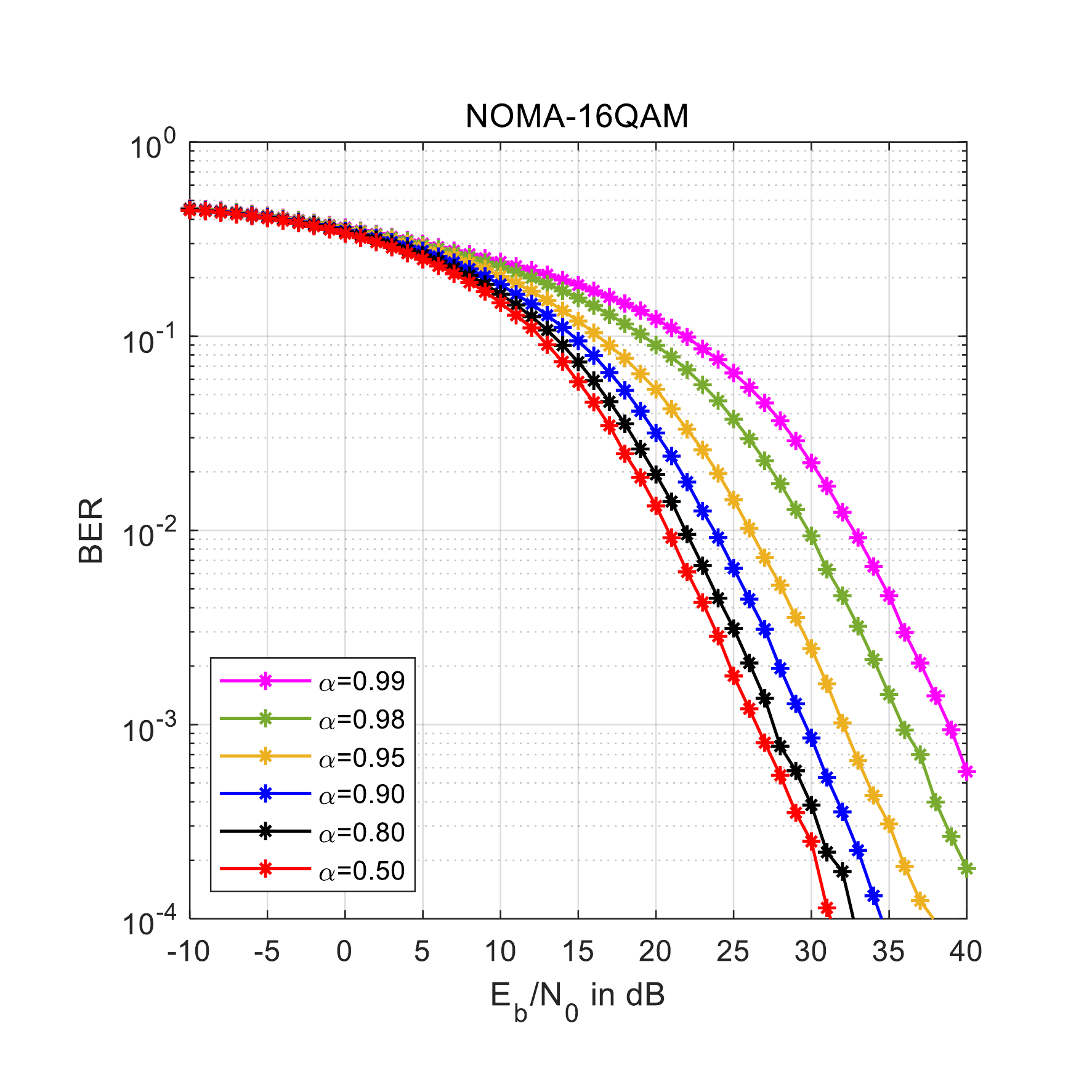}
  \caption{BER performance of PD-NOMA with 16QAM and different values of the power imbalance.}
  \label{Fig:16QAM}
\end{figure}

The simulation results are reported in Fig. \ref{Fig:QPSK} for QPSK and in Fig. \ref{Fig:16QAM} for 16QAM. As can be seen in both figures, the best performance results are obtained with $\alpha=1/2$, and they degrade as this parameter is increased. With $\alpha=0.9$ corresponding to an amplitude imbalance of 9.5 dB, the SNR degradation with respect to the amplitude balanced case at the BER of $10^{-3}$ is about 4.5 dB in QPSK and 3 dB in 16QAM. With $\alpha=0.95$ corresponding to an amplitude imbalance of 12.8 dB, the degradation is about 7 dB in QPSK and 6 dB in 16QAM. With $\alpha=0.98$ corresponding to an amplitude imbalance of 16.9 dB, the degradation is increased to 11.5 dB in QPSK and 9.5 dB in 16QAM. Finally, with $\alpha=0.99$ corresponding to an amplitude imbalance of 19.95 dB, the degradation is as high as 14 dB in QPSK and 12.5 dB in 16QAM. These results confirm the theoretical finding of the previous section that the best performance in memoryless NOMA is achieved when the user signals have perfect power balance at the receiver. This is actually what Multi-User MIMO (or Virtual MIMO) with power control does, and this technique became popular in wireless networks almost a decade before the surge of interest in Power-domain NOMA. Also note that the QPSK results of Fig. \ref{Fig:QPSK} and those of Table \ref{TABLE1} confirm that the ABEP upper bound given by eqn. (\ref{eq4}) is very tight. Indeed, the upper bound derived from Table \ref{TABLE1} reads a BER of $8\times10^{-4}$ for $\alpha=0.5$ and $5\times10^{-3}$ for $\alpha=0.9$, and the simulation results with $E_b/N_0$ = 20 dB read a BER of $5\times10^{-4}$ for $\alpha=0.5$ and $3\times10^{-3}$ for $\alpha=0.9$.

\section{Conclusions}
\label{sec5}
This paper analyzed the power imbalance factor between user signals on the uplink of a 2-user Power-domain NOMA system assuming uncorrelated Rayleigh fading channels and a maximum-likelihood receiver for signal detection. The results revealed that for a given total transmit power by the users, the minimum value of the average bit error probability is achieved when the power imbalance is zero, i.e., when the powers received from the two users are identical, as in Multi-User MIMO with perfect power control. This finding leads to the questioning the PD-NOMA principle when the channels are uncorrelated. The principle of transmitting in the same frequency band and at the same time instants a strong user signal and a weak user signal along with their detection using a SIC receiver has been long used to demonstrate that non-orthogonal multiple access leads to a higher capacity than orthogonal multiple access. The research community later used this principle to design NOMA without questioning whether the power imbalance is needed. Our analysis has shown that it is actually better from an average bit error rate point of view to have a perfect balance between the received signal powers. Our study was made for a 2-user NOMA uplink and assuming that the same signal constellation is used by both users. Extension to more than 2 users and also to NOMA systems where the users employ different constellations remain as topics for further studies.


\begin{thebibliography}{99}
\bibitem{YS}
Y. Saito, Y. Kishiyama, A. Benjebbour, T. Nakamura, A. Li, and K. Higuchi, ``Non-orthogonal multiple access (NOMA) for cellular future radio access,'' in \emph{IEEE 77th Vehicular Technology Conference (VTC Spring)}, Dresden, Germany, June 2-5, 2013, pp. 1--5.

\bibitem{ZD}
Z. Ding, Z. Yang, P. Fan, and H. V. Poor, ``On the performance of non-orthogonal multiple access in 5G systems with randomly deployed users,'' \emph{IEEE Signal Processing Letters}, vol. 21, no. 12, pp. 1501--1505, Dec. 2014.

\bibitem{LD}
L. Dai, B. Wang, Y. Yuan, S. Han, C.-L. I, and Z. Wang, ``Non-orthogonal multiple access for 5G: solutions, challenges, opportunities, and future research trends,'' \emph{IEEE Communications Magazine}, vol. 53, no. 9, pp. 74--81, Sept. 2015.

\bibitem{ZDING}
Z. Ding, Y. Liu, J. Choi, Q. Sun, M. Elkashlan, C.-L. I, and H. V. Poor, ``Application of non-orthogonal multiple access in LTE and 5G networks,'' \emph{IEEE Communications Magazine}, vol. 55, no. 2, pp. 185--191, Feb. 2017.

\bibitem{XLEI}
Z. Ding, X. Lei, G. K. Karagiannidis, R. Schober, J. Yuan, and V. Bhargava, ``A survey on non-orthogonal multiple access for 5G cellular networks: research challenges and future trends,'' \emph{IEEE Journal on Selected Areas in Communications}, vol. 35, no. 10, Oct. 2017.

\bibitem{SY}
S. Yang, P. Chen, L. Liang, J. Zhu, and X. She, ``Uplink multiple access schemes for 5G: a survey,'' \emph{ZTE Communications}, vol.15, no. S1, pp. 31--40, June. 2017.

\bibitem{MS}
M. Shirvanimoghaddam, M. Dohler, and S. J. Johnson, ``Massive non-orthogonal multiple access for cellular IoT: potentials and limitations,'' \emph{IEEE Communications Magazine}, vol. 55, no. 9, pp. 55--61, Sept. 2017.

\bibitem{HSLETTER}
H. Sari, F. Vanhaverbeke, and M. Moeneclaey, ``Multiple access using two sets of orthogonal signal waveforms,'' \emph{IEEE Communications Letters}, vol. 4, no. 1, pp. 4--6, Jan. 2000.

\bibitem{HSMAGAZINE}
H. Sari, F. Vanhaverbeke, and M. Moeneclaey, ``Extending the capacity of multiple access channels,'' \emph{IEEE Communications Magazine}, vol. 38, no. 1, pp. 74--82, January. 2000.

\bibitem{HSA}
H. Sari, A. Maatouk, E. Caliskan, M. Assaad, M. Koca, and G. Gui, ``On the foundation of NOMA and its application to 5G cellular networks,'' in \emph{IEEE Wireless Communications and Networking Conference (WCNC 2018)}, April 15-18, 2018, Barcelona, Spain, pp. 1--6.

\bibitem{AME}
A. Maatouk, E. Caliskan, M. Koca, M. Assaad, G. Gui, and H. Sari, ``Frequency-domain NOMA with two sets of orthogonal signal waveforms,'' \emph{IEEE Communications Letters}, vol. 22, no. 5, pp. 906--909, May. 2018.

\bibitem{AAX}
A. Al Khansa, X. Chen, Y. Yin, G. Gui, and H. Sari, ``Performance analysis of power-domain NOMA and NOMA-2000 on AWGN and Rayleigh fading channels,'' \emph{Physical Communication}, vol. 43, article number: 101185, Dec. 2020.

\bibitem{IC}
I. Cosandal, M. Koca, E. Biglieri, and H. Sari, ``NOMA-2000 vs. PD-NOMA: an outage probability comparison,'' \emph{IEEE Communications Letters}, vol. 25, no. 2, pp. 427--431, Feb. 2021.

\bibitem{MV}
M. Vaezi, Z. Ding, and H. V. Poor, ``Multiple access techniques for 5G wireless networks and beyond,'' \emph{Springer}, 2019.

\bibitem{QH}
Q. H. Spencer, C. B. Peel, A. L. Swindlehurst, and M. Haardt, ``An introduction to the multi-user MIMO downlink,'' \emph{IEEE Communications Magazine}, vol. 42, no. 10, pp. 60--67, Oct. 2004.

\bibitem{AM}
A. Mezghani, M. Joham, R. Hunger, and W. Utschick, ``Transceiver design for multi-user MIMO systems,'' in \emph{International ITG/IEEE Workshop on Smart Antennas (WSA 2006)}, Ulm, Germany, March 13--14, 2006, pp. 1--8.

\bibitem{SK}
S. K. Jayaweera, ``Virtual MIMO-based cooperative communication for energy-constrained wireless sensor network,'' \emph{IEEE Transactions on Wireless Communications}, vol. 5, no. 5, pp. 984--989, May 2006.

\bibitem{XC}
X. Chen, H. Hu, H. Wang, H.-H. Chen, and M. Guizani, ``Double proportional fair user pairing algorithm for uplink virtual MIMO systems,'' \emph{IEEE Transactions on Wireless Communications}, vol. 7, no. 7, pp. 2425--2429, July 2008.

\bibitem{BF}
B. Fan, W. Wang, Y. Lin, L. Huang, and K. Zheng, ``Spatial multi-user pairing for uplink virtual-MIMO systems with linear receiver,'' in \emph{IEEE Wireless Communications and Networking Conference (WCNC 2009)}, April 5-8 2009, Budapest, Hungary, pp. 1--5.

\bibitem{DT}
D. Tse and P. Viswanath, ``Fundamentals of wireless communications,'' \emph{Cambridge University Press}, 2005.

\bibitem{JS}
J. S. Yeom, H. S. Jang, K. S. Ko, and B. C. Jung, ``BER performance of uplink NOMA with joint maximum-likelihood detector,'' \emph{IEEE Transactions on Vehicular Technology}, vol. 68, no. 10, pp. 10295--10300, Oct. 2019.

\bibitem{HS}
H. Semira and F. Kara, ``Error performance of uplink SIMO-NOMA with joint maximum-likelihood and adaptive M-PSK,'' \emph{Proc. 2021 IEEE Int. Black Sea Conf. on Communications and Networking (Virtual Conference)}, May 2021.

\end{thebibliography}
\end{document}